\documentclass[aps,prb,showpacs,twocolumn,superscriptaddress,amsmath,amssymb,floatfix]{revtex4-1}
\usepackage{tabularx,wasysym}
\usepackage{bm}
\usepackage{epsfig,subfigure}
\usepackage{multirow}
\usepackage{graphicx}
\usepackage{color}
\usepackage{amsfonts}
\usepackage{exscale}
\usepackage{amsbsy}
\graphicspath{{Fig/}}
\usepackage[normalem]{ulem} 
\usepackage{soul}
\usepackage[colorlinks,urlcolor=blue,linkcolor=blue,anchorcolor=blue,citecolor=blue]{hyperref}

\begin{document}

\title{Shadow of complex fixed point: Approximate conformality of Q $>$ 4 Potts model
}
\author{Han Ma}
\affiliation{Department of Physics, University of Colorado, Boulder, Colorado 80309, USA}
\author{Yin-Chen He}
\affiliation{Perimeter Institute for Theoretical Physics, Waterloo, Ontario N2L 2Y5, Canada}

\date{\today}

\begin{abstract}
We study the famous example of weakly first order phase transitions in the 1+1D quantum $Q$-state Potts model at $Q>4$. 
We numerically show that these weakly first order transitions have approximately conformal invariance.
Specifically, we find entanglement entropy on considerably large system sizes fits perfectly with the universal scaling law of this quantity in the conformal field theories (CFTs). This supports that the weakly first order transitions is proximate to complex fixed points, which are described by recent conjectured complex CFTs. 
Moreover, the central charge extracted from this fitting is close to the real part of the complex central charge of these complex CFTs. 
We also study the conformal towers and the drifting behaviors of these conformal data (e.g., central charge and scaling dimensions).
\end{abstract}
\maketitle

\section{Introduction} 

The study of phase transitions in strongly correlated systems is constantly contributing to the discovery of new physics~\cite{cardy1996scaling, sachdev2011quantum}. 
The most studied examples are continuous phase transitions since they have scale invariance indicating universal scaling behaviors~\cite{cardy1996scaling}. 
They also correspond to renormalization group (RG) fixed points (FP) described by conformal field theories (CFTs)~\cite{francesco2012conformal}, if there is Lorentz invariance. 
On the contrary, there are conceptually distinct first order phase transitions~\cite{fisher1982scaling,binder1987theory} whose importance are significantly overlooked due to the lack of universal behavior. But it is time to end this situation because a new physical mechanism is discovered for a certain type of weakly first order phase transitions, whose weakness can be attributed to the existence of nearby complex fixed points (cFP)~\cite{kaplan2009conformality, wang2017dqcp, gorbenko2018walking,gorbenko2018walking2}. 

\begin{figure}
\includegraphics[width=.49\textwidth]{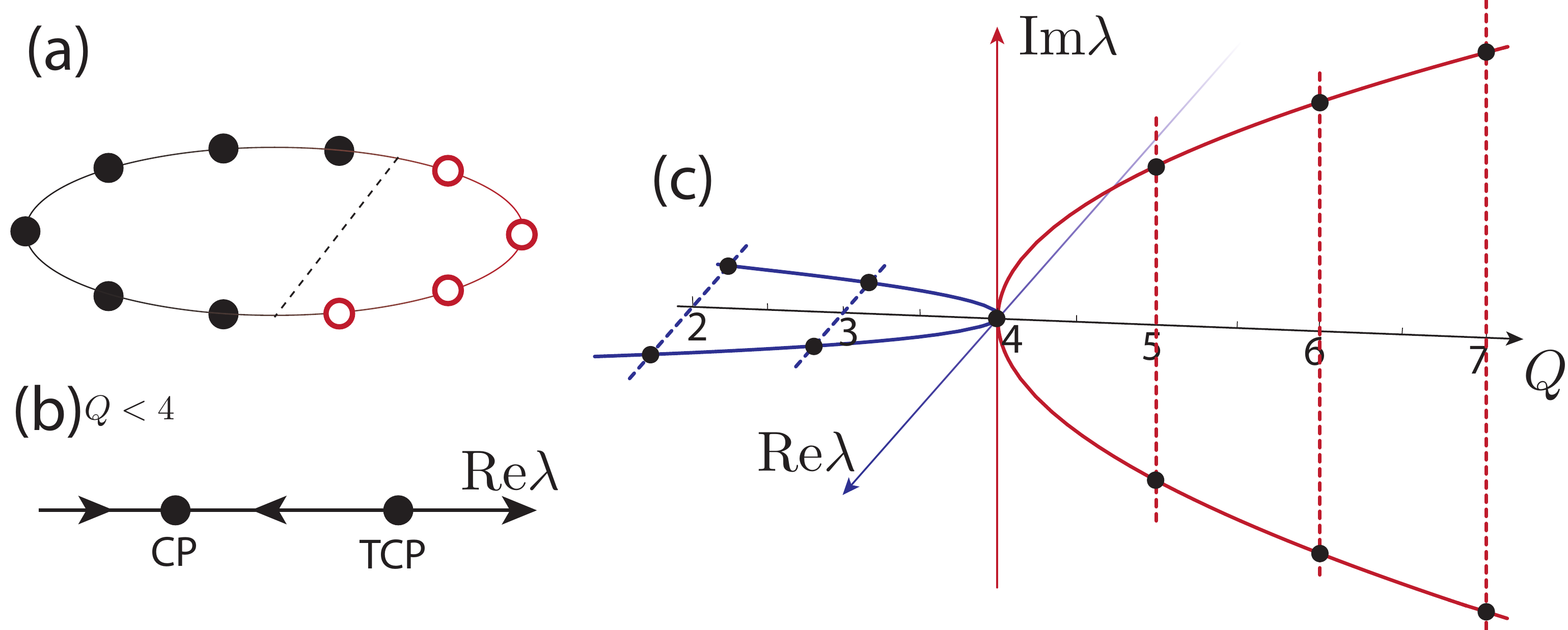}
\caption{(a) A bi-partition of a periodic Potts chain. (b) The critical point (CP) and tricritical point (TCP) when $Q<4$. $\lambda$ is irelevant at critical point and relevant at the tricritical point. (c) FPs at different $Q$. \label{fig:illustration}}
\end{figure}

One famous example of such a weakly first order phase transition occurs in a 2D statistical mechanical model called the  $Q$-state Potts model~\cite{wu1982potts} with $Q=5$.
Equivalently, we can define a quantum Potts chain (Fig.~\ref{fig:illustration}(a)) with Hamiltonian
\begin{equation}
H^{\rm Potts} = - \sum_{i=1}^L \sum_{k=1}^{Q-1} (h \Omega_i^k + J M_i^k M_{i+1}^{Q-k}).\label{eq:ham_potts}
\end{equation}
At each site, there are $Q$ possible spin states denoted as $|  m_i  \rangle $, where $m_i = 0, \dots, Q-1$. 
Then, we have $\Omega_i |  m_i \rangle = |  (m_i +1) \textrm{ mod } Q \rangle$ and $M_i |  m_i  \rangle  = e^{i2\pi m_i/Q} |  m_i \rangle $. 
Clearly for $Q=2$, this model reduces to the transverse field Ising model. 
For any integer $Q$, the model has two possible phases: 1) an ordered phase spontaneously breaking the $S_Q$ symmetry at $J>h$; 2) a disordered phase at $J<h$. The Hamiltonian is self-dual at the transition point $J=h$, where the transition is continuous for $Q \leq 4$ and is of first order for $Q>4$. At $Q=5$, the phase transition exhibits extremely weakly first order behavior, namely it has a huge correlation length~\cite{buddenoir1993correlation, delfino2000field} even beyond computing power~\cite{iino2018detecting}.
Surprisingly, this weakly first order transition results from a completely new physics as discussed in Ref.~\onlinecite{wang2017dqcp,gorbenko2018walking2}.

If we generalize the $Q$-state quantum Potts model by including perturbations of symmetry allowed singlet operators~\cite{wu1982potts}, then the changeover from continuous to first order transition of the original model can be understood through RG equations of couplings of low-lying singlets. Enforcing self-duality, the coupling of first relevant singlet $\varepsilon$, i.e., tuning operator for the order-disorder transition, is set to be zero.
Now we consider the RG equation for the coupling $\lambda$ of the subleading singlet operator, $\varepsilon'$~\cite{cardy1980scaling}:
\begin{equation}
-\frac{d\lambda}{d \ln L} =  a (4-Q) -  \lambda^2 \label{eq:marginal}
\end{equation}
where $a>0$ is a real constant. When $Q < 4$, there are two real fixed points (rFP) corresponding to a critical point (attractive) and a tricritical point (repulsive) at real $\lambda$, as shown in Fig.~\ref{fig:illustration}(b). 
The critical point corresponds to the continuous transition between ordered and disordered phases in the original Potts model. 
The two rFPs collide at $Q=4$ and  
disappear when $Q>4$.  
But if we extend $\lambda$ to be a complex coupling, there are two cFPs~\cite{wang2017dqcp} at $\lambda_\pm = \pm i \sqrt{a|Q-4|}$ when $Q>4$. 
The creation of cFPs by the collision of two rFPs is illustrated in Fig.~\ref{fig:illustration}(c). 
Although without encountering any rFP, the RG flow of the real coupling $\lambda$ between two cFPs is extremely slow for small $Q-4$, leading to a weakly first order phase transition with approximate conformality ~\cite{gorbenko2018walking, gorbenko2018walking2}. 
Therefore, the weakness of this first order phase transition can be attributed to the existence of cFPs and it does not rely on any fine tuning of the physical couplings or interactions.

More appealingly, it was proposed that the cFPs are described by complex CFTs~\cite{gorbenko2018walking, gorbenko2018walking2}. 
They are non-unitary CFTs with complex central charges and scaling dimensions, in contrast to other known non-unitary CFTs, e.g., Lee-Yang singularity~\cite{lee1952yang, cardy1985conformal} and percolation problem~\cite{cardy1992critical, cardy2001conformal}, whose conformal data are all real. 
We note that, cFPs and complex scaling dimensions have been noticed and discussed in other quantum field theories for a decade~\cite{dymarsky2005perturbative,kaveh2005chiral,gies2006chiral, kaplan2009conformality,braun2014phase,giombi2017bosonic,giombi2018prismatic,klebanov2018tasi}.
The complex conformal data of $Q>4$ Potts model can be obtained by the analytically continued partition function of Coulomb gas, an alternative description of the generalized Potts model~\cite{di1987relations}.
As elaborated before, the subleading singlet operator $\varepsilon'$ plays an important role in the appearance of the cFPs. 
When $Q<4$, $\varepsilon'$ is irrelevant at the critical point, but relevant at the tricritical point. 
At $Q=4$, the two FPs coincide, and $\varepsilon'$ is exactly marginal.
When $Q>4$, the two FPs becomes complex, yielding complex scaling dimensions $\Delta_{\varepsilon'}$ for $\varepsilon'$.
Interestingly, we find that $\Delta_{\varepsilon'}$ still has modulus exactly equal to $2$, i.e. $\sqrt{({\rm Re} \Delta_{\varepsilon'})^2 + ({\rm Im} \Delta_{\varepsilon'})^2} =2$ (see Appendix ~\ref{app:tower}).
It might be a coincidence, or it corresponds to certain type of marginality in the complex CFTs of $Q>4$ Potts model that could hold for other complex CFTs.

A natural question arising from this proposal of complex CFTs is whether we can observe the approximate conformality in the weakly first order transition point of the $Q>4$ Potts model.
In this letter, we give a definite answer to this question.
We numerically calculate the entanglement entropy (EE) at the transition point, and we find it scales perfectly with the scaling formula of a true CFT~\cite{vidal2003entanglement,calabrese2009entanglement}. 
Accordingly, we read off the central charge and find its value is close to the real part of complex central charge of the conjectured complex CFTs~\cite{gorbenko2018walking2}. 
However, the conformal invariance is only approximate, as the central charge is drifting with the system size. 
We study this drift behavior in detail and find it is consistent with our theoretical expectation. 
We also study the conformal tower at the transition point as well as their drift behavior.
Our discussion of quantum Potts model may also have interesting implications beyond previous discussions of classical Potts model~\cite{gorbenko2018walking2}.

\section{Central charge $c$}
It is well known that the EE of a CFT follows a universal scaling depending on the CFT's central charge~\cite{vidal2003entanglement, calabrese2009entanglement}. 
Specifically, we consider A-B bi-partition on a length-$L$ periodic chain (shown in Fig.~\ref{fig:illustration}(a)), and calculate EE as $S= \textrm{Tr} \rho_A \ln \rho_A $ through the reduced density matrix $\rho_A=\textrm{Tr}_B (|\psi\rangle \langle \psi|)$ of ground state $|\psi \rangle$.
For a CFT, EE scales universally with the subsystem size ($X$),
\begin{equation}
S(X)= \frac{c}{3} \ln \Big( \frac{L}{\pi}\sin \frac{\pi X}{L} \Big) + S_0,
 \label{eq:entropy_scale}
\end{equation}
with $c$ being the central charge.

\begin{figure}[h]
\includegraphics[width=.38\textwidth]{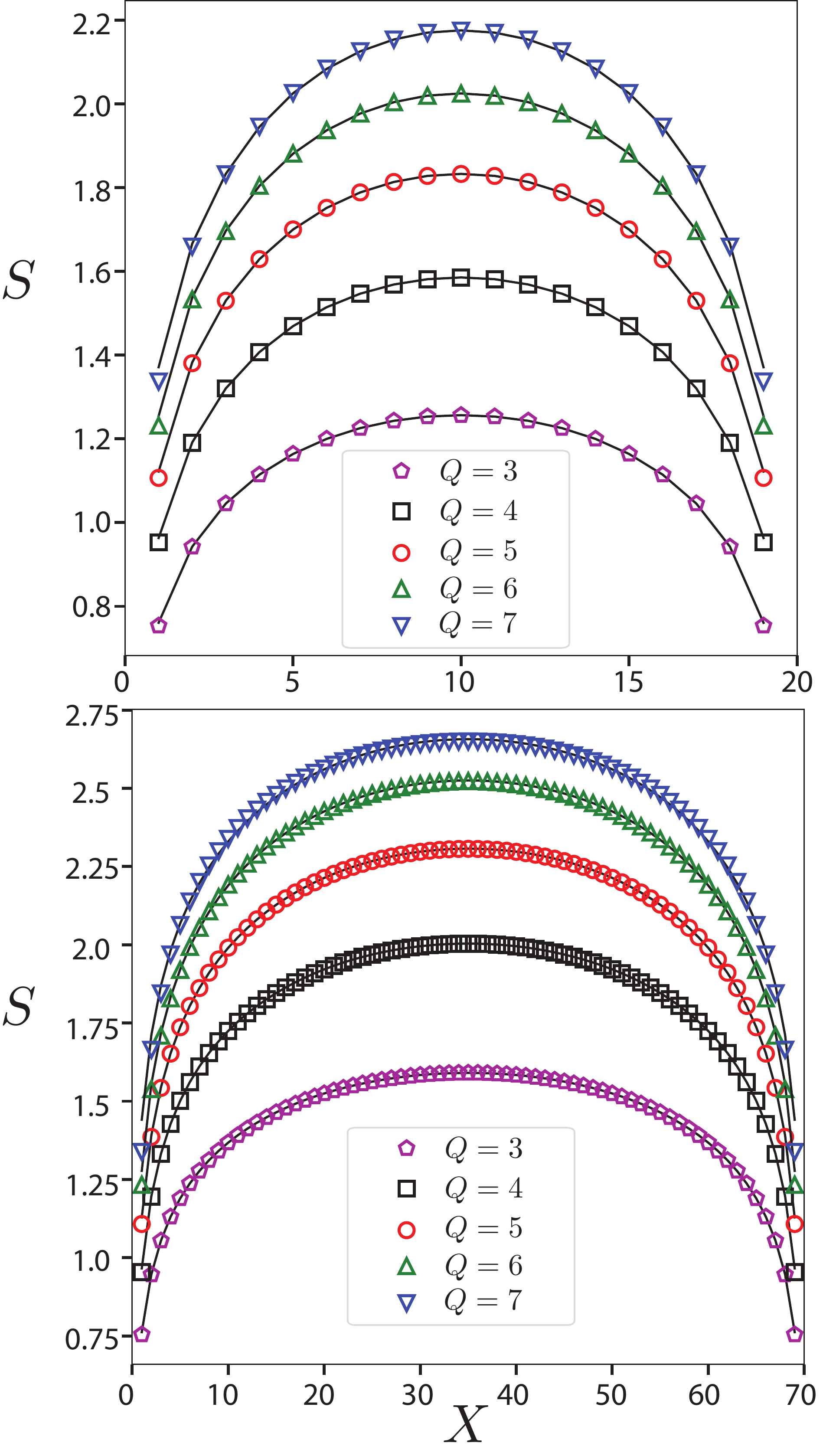}
\caption{Scaling of the EE ($S$) at the transition point of $Q$-state quantum Potts model for different system sizes $L$. We fit the data with function in Eq.~\eqref{eq:entropy_scale}. The fitted central charges are listed in Table~\ref{tab:central_charge}. \label{fig:central_charge}}
\end{figure}

Numerically, we use density matrix renormalization group (DMRG)~\cite{white1993density} to calculate the $Q$-state Potts model on a periodic chain.
Fig.~\ref{fig:central_charge} plots the EE as functions of $X$ for $Q=3,4,5,6,7$, which are fitted with Eq.~\eqref{eq:entropy_scale}.
For all the $Q$s we consider, the EE fits almost perfectly with the CFT's entropy formula.
Indeed, from these numerical data, it is hard to tell the difference between the continuous transition at $Q\le 4$ and the weakly first order transition at $Q>4$.
This strongly suggests that, even though being first order phase transitions at $Q>4$, they still have approximate conformality. 

The fitted central charge $c$ is summarized in Table~\ref{tab:central_charge}.
For $Q=3$ and $Q=4$, we get central charge $c=4/5$ and $c=1$ as expected for the corresponding CFTs, namely the $\mathcal{M}_{6,5}$ minimal model~\cite{francesco2012conformal} and the $\mathbb{Z}_2$ orbifold of the $U(1)_8$ CFT (equivalently the $\mathbb{D}_2$ orbifold of the $SU(2)_1$ CFT)~\cite{dijkgraaf1989operator}. 
For $Q>4$, $c$ from the EE fitting is close to the real part of complex central charge $\textrm{Re}(c_{\rm exact})$ of the proposed complex CFTs which has $Q$ dependence as~\cite{gorbenko2018walking2, di1987relations}
\begin{equation}
c^\pm_{\rm exact}=1+\frac{3}{\pi} \frac{ \left[ \cosh^{-1}\Big(\frac{Q-2}{2}\Big) \right]^2}{2 \pi \mp i \cosh^{-1}\Big(\frac{Q-2}{2}\Big) } .
\end{equation}
The exact values for particular $Q$s are listed in Table~\ref{tab:central_charge}.
As system size increases, we find $c$ approaches to $\textrm{Re}(c_{\rm exact})$ for $Q=5$. On the other hand, when $Q=6$ or $Q=7$, $c$ deviates from ${\rm Re}(c_{\rm exact})$ as system size increases. This means the central charge $c$ drifts as a function of length scale. Thus, the approximately conformal invariance doesn't give a size independent central charge, instead it follows a universal drift behavior, which can be understood analytically~\cite{gorbenko2018walking2}. 
It can also be regarded as a particular type of finite size effect. 
Below, we will study this drift behavior in more detail.

\begin{table}[h]
\begin{tabular}{c|c|c|c|c|c} \hline
Q & 3 & 4 & 5 & 6 & 7 \\ \hline
$\textrm{Re}(c_{\rm exact})$ & $0.8$ & $1$ & $1.13755$ & $1.2525 $ & $1.35125 $ \\
$| \textrm{Im}(c_{\rm exact}) |$ & $0$ & $0$ & $0.02107$ & $0.05292$ &$0.08759$ \\
$c_{L=20}$ & 0.802 & 1.006 &  1.149 & 1.246 & 1.301 \\
$c_{L=70}$ & 0.800 & 1.002 & 1.139 & 1.205 & 1.175 \\
$c_{R}$ & - & - &  1.143 & 1.244 & 1.366 \\
\hline
\end{tabular}
\caption{Central charges by fitting Eq.~\eqref{eq:entropy_scale} and Eq.~\eqref{eq:central_charge_drift} for $Q$-state quantum Potts model with different system sizes. These values are close to real part of the exact central charges at cFPs. \label{tab:central_charge}}
\end{table}

\section{Central charge drift}
We start with the celebrated c-theorem~\cite{zamolodchikov1986irreversibility}, which states how central charge changes with the RG coupling,
\begin{equation}
-\frac{d \lambda}{d \ln L} = \frac{\partial}{\partial \lambda} c(\lambda). \label{eq:c-theorem}
\end{equation}
Here $\lambda$ is a coupling constant which runs under RG flow (changing of system size $L$).
In the UV (i.e., small system size), the measured central charge $c\left[ \lambda(L) \right]$ has a strong size dependence.
As the FP (at $\lambda_c$) is approached, we will have $\lim_{\lambda(L) \rightarrow \lambda_c} c\left[ \lambda(L) \right] = c_{\rm exact}$, independent of scale $L$.

Together with Eq.~\eqref{eq:marginal} and Eq.~\eqref{eq:c-theorem}, we are able to obtain the drift behavior of central charge $c_{\rm drift}(L)$ as a function of length scale $L$:
\begin{equation}
c_{\rm drift}(L) = c_{R} - \alpha \tan \Big(\gamma \ln \frac{L}{L_0} \Big)
+ \dots \label{eq:central_charge_drift}
\end{equation}
where $L$ is the length scale, which is the system size in our discussion. 
$L_0$ is a length scale where $\lambda(L_0)=0$. 
This formula applies when the correction to $c_R$ is small, approximately $1/10 \lessapprox L/L_0 \lessapprox 10 $, namely the RG flow is close to the cFPs.
This form of drift is quite generic~\cite{dymarsky2005perturbative, nogueira2013deconfined, gorbenko2018walking2}, although $\alpha$, $\gamma$ and $L_0$ are model dependent quantities. For $Q$-state Potts model, we have $\alpha=\left[a(Q-4)\right]^{3/2}$ and $\gamma= \sqrt{a(Q-4)}$ with $a=1/\pi^2$ at one loop order.
$c_R$ is the real part of the central charge of the complex CFTs.
Appendix~\ref{app:c_drift} gives a detailed derivation of Eq.~\eqref{eq:central_charge_drift}.

\begin{figure}[h]
\includegraphics[width=.5\textwidth]{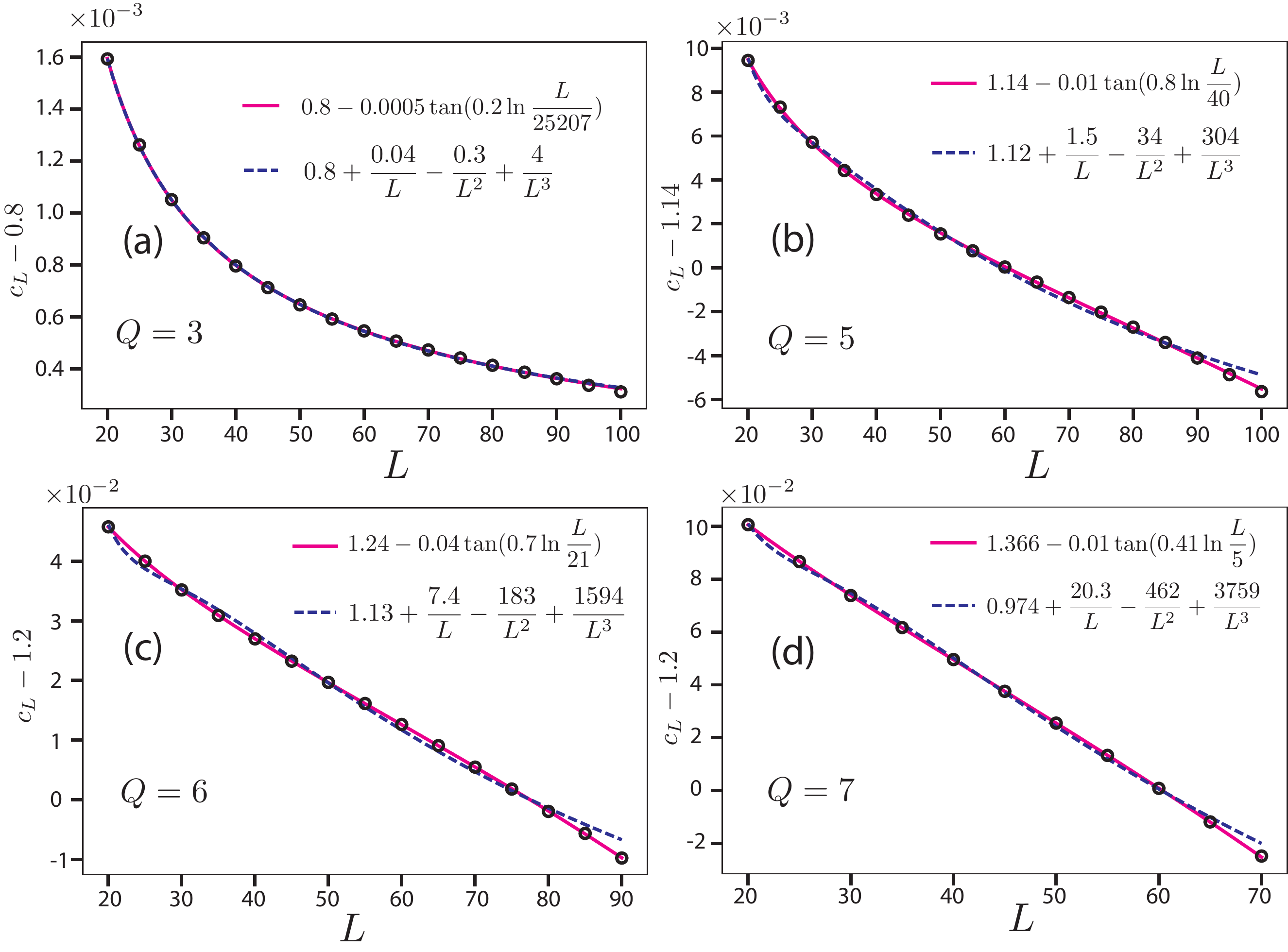}
\caption{The central charge at transition point for $Q=3,5,6,7$ as a function of system size is fitted with formula in Eq.~\eqref{eq:central_charge_drift}. 
As a comparison, it is also fitted with usual polynomial of $1/L$.
\label{fig:c_drift}}
\end{figure}

Numerically, we obtain the central charges at $Q=3,4,5,6,7$ for different system sizes. These data are shown as circles in Fig.~\ref{fig:c_drift}. 
The central charges of different system sizes $L$ can be fitted with the drift formula in Eq.~\eqref{eq:central_charge_drift}, with all the parameters $c_{R}$, $a$, $b$, and $L_0$ to be fitted. 
In comparison, we also fit the data with a polynomial of $1/L$: $c_{\rm poly} (L) = c_p + \frac{\alpha}{L} + \frac{\gamma}{L^2} + \frac{\eta}{L^3} + \dots$,
a usual form of finite size scaling. 

In Fig.~\ref{fig:c_drift}(a), the polynomial fits the data of $Q=3$ pretty well, agrees with the fact that the system flows into a CFT. 
One may note that Eq.~\eqref{eq:central_charge_drift} also fits the data well. 
However, the fitting coefficient $a$ is extremely small and $L_0$ is extremely large, resulting in approximate power law behavior. 
Similar behavior is also observed for $Q=4$, but with a stronger finite size effect (see Appendix~\ref{app:c_q=4}). 

On the contrary, when $Q>4$, the drift formula Eq.~\eqref{eq:central_charge_drift} fits much better than the polynomial one, as shown in Fig.~\ref{fig:c_drift}(b)-(d). This means central charge obeys the drift behavior described by Eq.~\eqref{eq:central_charge_drift}. 
$c_R$ obtained by this fitting is close to the theoretical value, but $\alpha$ is a  few times larger than our one-loop calculation. 
Moreover, $\gamma$ decreases as $Q$ goes up, distinct from theoretical expectation.
This discrepancy may be due to 
scale-dependent deviations from (space-time) rotational invariance, parametrized by `running of the speed of light', in a quantum phase transition~\cite{gorbenko2018walking2}. 
Nevertheless, as $Q$ increases from $5$ to $7$, $\alpha$ increases gradually, indicating the drift effect is getting stronger. 
This is consistent with another observation that $L_0$ decrease as $Q$ increases. 
They provide supportive evidence for the fact that the cFPs are moving further away from the real coupling $\lambda$ as $Q$ increases. 
This naturally leads to a faster RG flow and hence a stronger violation of conformal invariance at larger $Q$. 

\section{Conformal tower}

In this section, we numerically extract the conformal tower (i.e., scaling dimensions of operators) of the Potts chain at the phase transition point. 
As is well known, the conformal tower can be obtained via the state-operator correspondence of radial quantization of $1+1$D CFTs on a $S^1\times R$ manifold.
For example, one can interpret $S^1$ as the space dimension, $R$ as the imaginary time dimension.
Then the energies of excited states (scaled with system size) of corresponding quantum Hamiltonian $H_{\textrm{CFT}}$ are nothing but the scaling dimensions of CFT operators (Eq.~\eqref{eq:energy_scaling} below)~\cite{affleck1988universal, blote1986conformal,  zou2017conformal, milsted2017extraction}. 

\begin{figure}[h]
\includegraphics[width=.49\textwidth]{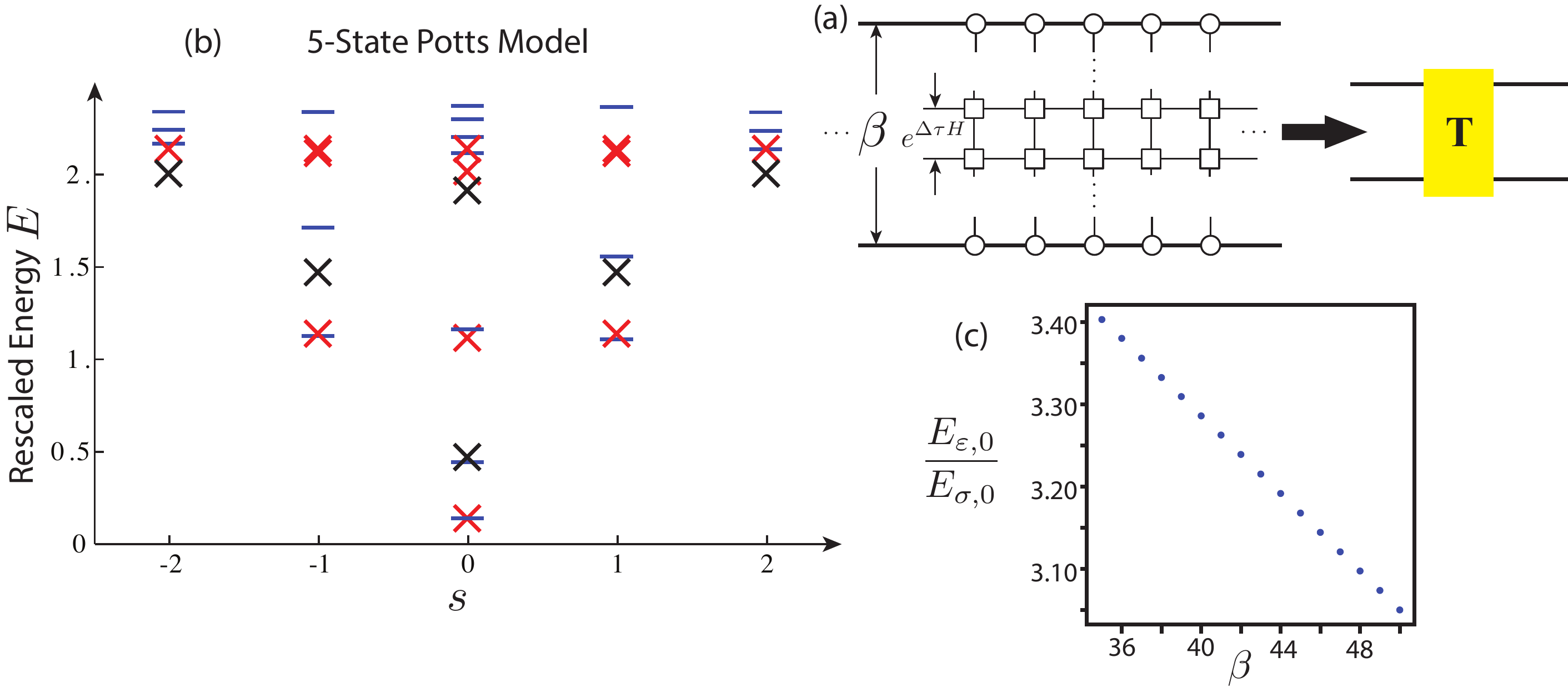}
\caption{(a) Schematic illustration of tensor network calculation of a thermal partition function. The tensors denoted as circles are identified due to the periodic boundary condition along $\beta$. (b) Conformal tower of the 5-State quantum Potts model. Crosses (Red: primaries; Black: descendants) are theoretical values while lines are numerical results. The spectrum is separated into sectors with different conformal spin (s).
(c) The drift of the ratio of scaling dimensions $\Delta_{\varepsilon}/\Delta_\sigma$. \label{fig:drift}}
\end{figure}

Alternatively, one can do a space-time rotation and consider a thermal partition function of $e^{-\beta H_{\textrm{CFT}}}$ at a finite temperature $\beta=1/T$ ($S^1$, equivalently imaginary time) and infinite size in the spatial direction ($R$).
The energy of eigenstates $E_{\phi,n}$ can be calculated as the inverse of the temporal correlation length which is encoded in the eigenvalues of the temporal transfer matrix~\cite{schollwock2011density, tirrito2018characterizing}. The transfer matrix is defined in terms of temporal matrix product state as shown in Fig.~\ref{fig:drift}(a). Then, we have
\begin{equation}
E_{\phi,n} = \frac{2\pi v}{\beta}\Delta_{\phi,n} \label{eq:energy_scaling}
\end{equation}
for primary operator $\phi$ at level $n$, where $v$ is a non-universal velocity which might have size dependence on $\beta$ but is independent of $\phi$ and $n$.
$\Delta_{\phi,n}$ is its scaling dimension. $n\neq 0$ indicates descendants of $\phi$. The identity operator $\phi=\mathbf{1}$ corresponds to the ground state with $E_{\mathbf{1},0}=0$.
We note that $\beta$ defines the effective length scale in this scheme.

The conformal towers of $Q=2,3$ have been studied before (e.g., see Refs.~\onlinecite{li2015criticality,zou2017conformal, milsted2017extraction}). 
Here, we focus on the conformal tower of the 5-state Potts model. 
Since $v$ is unknown, we use the lowest level $E_{\sigma,0}$ as a reference, where $\sigma$ is the lowest primary operator in the vector representation of $S_Q$ symmetry. 
Specifically, we set $\Delta_{\sigma,0}$ equal to the real part of its theoretical value of the proposed complex CFTs.
In this way, $v$ can be obtained and the whole spectrum can be rescaled by this $v$. 
Fig.~\ref{fig:drift}(b) shows the conformal tower, it includes numerical data at $\beta=40$ (blue lines) and the theoretical values (crosses).
They match well at lower levels, while there is an obvious deviation for higher levels.

Furthermore, we plot the drift behavior of scaling dimensions~\cite{iino2018detecting,gorbenko2018walking2}. 
Since the velocity $v$ will also drift, we then consider the quantity $E_{\varepsilon,0}/E_{\sigma,0}=\Delta_{\varepsilon}/\Delta_\sigma$ with $\varepsilon$ being the lowest singlet operator (i.e., the second excited state). As shown in Fig.~\ref{fig:drift}(c), this quantity drifts significantly as the system size ($\beta$) changes. 

\section{Summary and Discussion}

In this work, we studied the weakly first order phase transitions in the $Q$-state quantum Potts chain with $Q>4$.
Based on the EE and conformal tower, we provided strong numerical evidence that these transitions have approximate conformality, which originates from the proximity to cFPs~\cite{wang2017dqcp,gorbenko2018walking,gorbenko2018walking2}. 
Particularly, we found EE perfectly follows the same universal scaling law of CFTs and the central charge extracted from it is close to the real part of the central charge of the proposed complex CFTs. 
We further discussed how these conformal data (i.e., central charge, scaling dimensions) drift with system size, which agrees with our theoretical expectation. 

Our current study supports the theoretical proposal of cFPs/complex CFTs in $Q>4$ Potts model, particularly in the quantum phase transition rather than the classical one discussed in Ref.~\onlinecite{gorbenko2018walking2}.
Moreover, the validity of approximate conformality in $Q>4$ quantum Potts model naturally leads to questions regarding its finite temperature physics.
Once temperature $T$ is higher than the tiny gap of weakly first order transition, the system may enter into the quantum critical fan controlled by the cFPs. 
It will be interesting to understand if this quantum critical fan is different from these of continuous quantum phase transitions~\cite{sachdev2011quantum}.
Finally, we remark that the cFP may also exist in other systems, particularly in higher dimensions.
One possible candidate is the deconfined phase transition between N\'eel state and valence bond solid in $2+1$ dimensions~\cite{senthil2004deconfined, senthil2004quantum, wang2017dqcp}, in which the drift of critical exponents was also observed numerically (e.g., see~\onlinecite{sandvik2007evidence, melko2008scaling, jiang2008antiferromagnet, lou2009antiferromagnetic, sandvik2010computational, harada2013possibility, chen2013deconfined}). 
We hope our current numerical study may help to resolve this long-standing puzzle.

\section*{Acknowledgement.}
We thank Igor Klebanov, Ashley Milsted, T. Senthil, Guifre Vidal, Yijian Zou for stimulating discussions.
We, in particular, are grateful to Chong Wang for discussions and collaborations at the early stage of this project.
This research was  enabled  in  part  by  support provided by Compute Canada (www.computecanada.ca). 
We have also used the TenPy tensor network library, which includes contributions from Mike Zaletel, Roger Mong, Frank Pollmann, and others. 
H.M. is supported by the grant of Michael Hermele from the U.S. Department of Energy, Office of Science, Basic Energy Sciences (BES) under Award No. DE-SC0014415.
Research at Perimeter Institute is supported by the Government of Canada through the Department of Innovation, Science and Economic Development Canada and by the Province of Ontario through the Ministry of Research, Innovation and Science. 

\bibliography{ref} 

\widetext
\appendix
\section{Derivation of central charge drift in Eq.~(6) \label{app:c_drift}}

The basic idea of this derivation is to combine the c-theorem and the RG flow of physical couplings. 

As discussed in Ref.~\cite{gorbenko2018walking2}, the physics of two cFPs as well as the nearby real axis is captured by the RG equation in Eq.~(2).
Specifically, this equation, together with the c-theorem in Eq.~(5), gives us the central charge depending on $\lambda$:
\begin{equation}
    c(\lambda)= c_0 + a(4-Q)\lambda - \frac{1}{3}\lambda^3
\end{equation}
At the FP ($\lambda=\lambda^\ast$), the central charge of corresponding (complex) CFT is
\begin{equation}
    c^\ast = c_0 +a(4-Q)\lambda^\ast - \frac{1}{3}(\lambda^\ast)^3
\end{equation}
When $Q>4$, $\lambda^\ast$ is pure imaginary. $c_0$ is thus the real part of the central charge $c_R$ at the cFP.
Furthermore, from Eq.~(2), we can obtain the length scale dependent coupling $\lambda$ when $Q>4$: 
\begin{equation}
    \lambda = \sqrt{a(Q-4)}\tan \left[\sqrt{a(Q-4)} \ln \frac{L}{L_0}\right] \label{eq:lambda_L}
\end{equation}
where $a=1/\pi^2$ for the Potts model and $L_0$ is the length scale at which $\lambda=0$. 

Finally, we can get the length scale dependent central charge:
\begin{equation}
    c(L) = c_R - \left[a(Q-4)\right]^{3/2}\tan \left[\sqrt{a(Q-4)}\ln \frac{L}{L_0}\right]
\end{equation}
when $\lambda \ll 1$. This is the Eq.~(6) in the main text.

\section{Central charge at $Q=4$ \label{app:c_q=4}}

\begin{figure}[h]
\includegraphics[width=.5\textwidth]{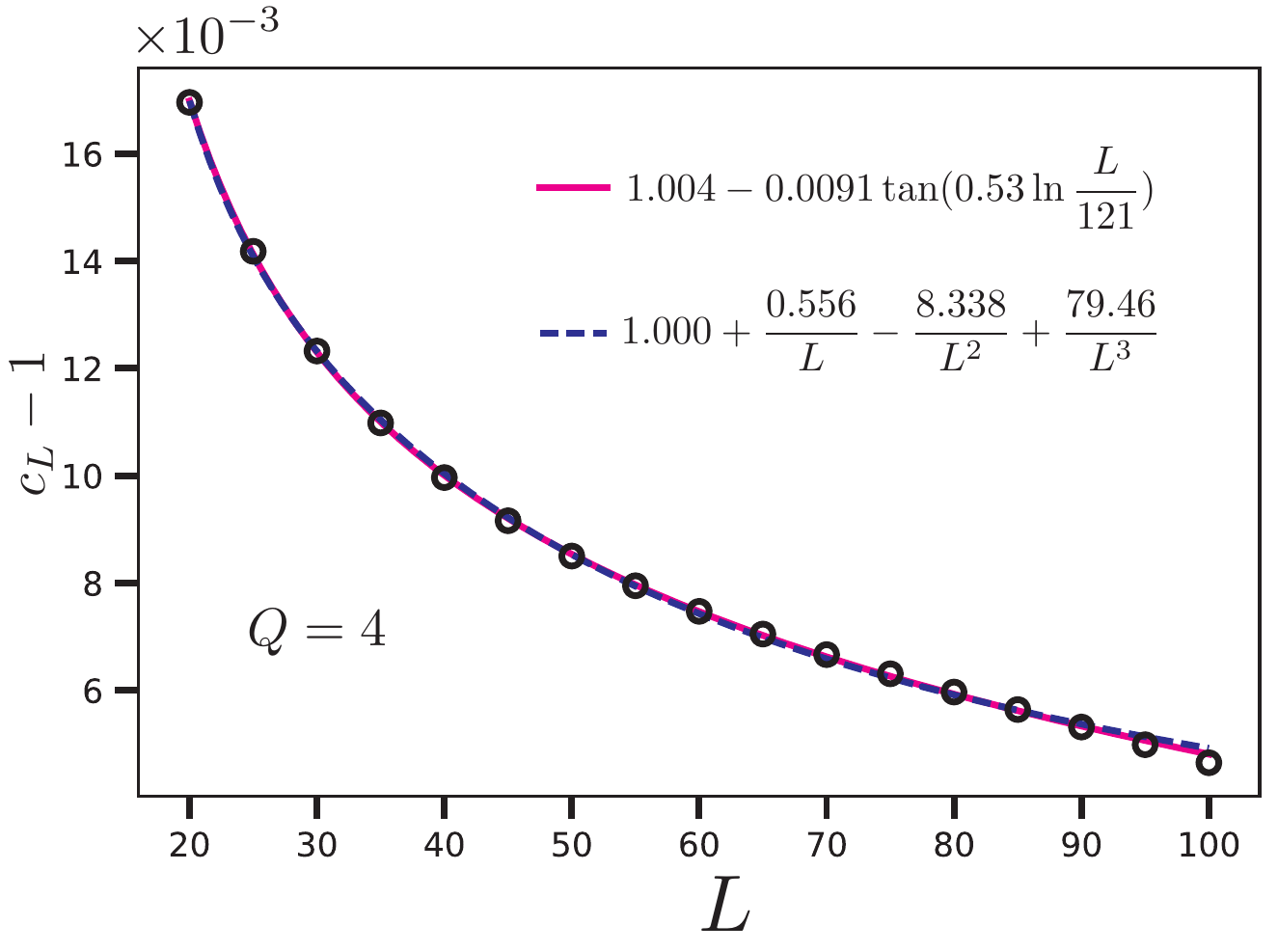}
\caption{The drift of central charge at $Q=4$ is fitted with the drift formula Eq.~(6) and the polynomial of $1/L$. \label{fig:c_drift_4}}
\end{figure}

Here we discuss central charge's dependence on system size $L$ for a true CFT at $Q=4$.
In this case, the two rFPs collide and there is an marginal operator $\varepsilon'$~\cite{dijkgraaf1989operator}. 
This leads to severe finite size effect. 
As shown in Fig.~\ref{fig:c_drift_4}, different from the $Q=3$ case, the power law does not fit well.
As a comparison, we also fit the central charge using 
the drift formula Eq.~(6). 
We also remark that central charge's size dependence is much smaller than the $Q=5, 6, 7$ Potts model.

\section{Conformal tower \label{app:tower}}

Firstly, we would like to mention that the scaling dimension of operator $\varepsilon'$ at general $Q \geq 4$ is
\begin{equation}
\Delta_{\varepsilon'} = \frac{8}{2- \frac{1}{\pi} \cos^{-1} \frac{Q-2}{2} }-2
\end{equation}
where $\cos^{-1} \frac{Q-2}{2} =i \cosh^{-1} \frac{Q-2}{2}$. One can directly verifies the complex scaling dimension of $\varepsilon'$ has modulus $2$ for general $Q \geq 4$.

\begin{figure}[h]
\includegraphics[width=.55\textwidth]{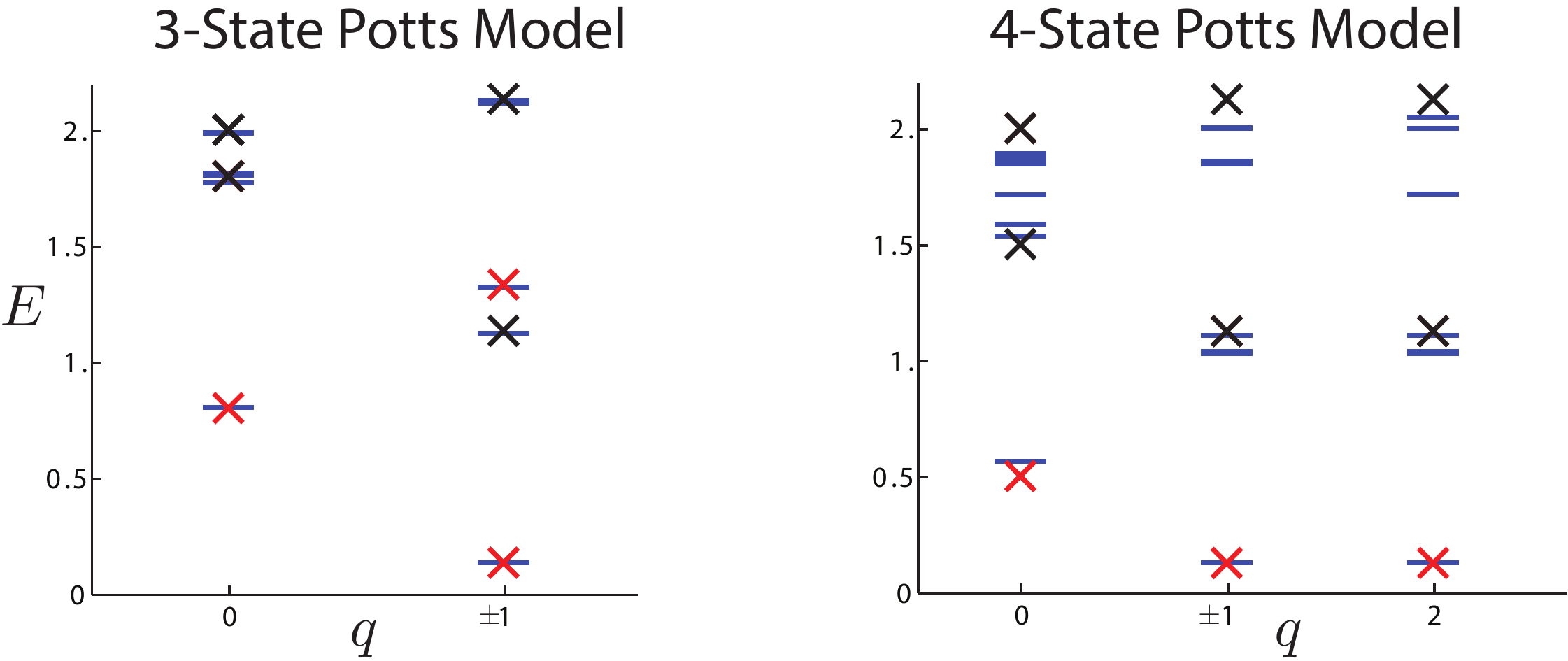}
\caption{Conformal tower of critical $Q=3$ ($\beta=40$) and $Q=4$ ($\beta=50$) Potts model. Red (primary) and Black (descendant) crosses are theoretical values while lines are numerical results. The spectrum is separated into sectors with different symmetry charge (q).\label{fig:tower_34}} 
\end{figure}

Next, we would like to present the numerical data of conformal tower at the transition point of the Potts chain. The results for $Q=3,4$ are shown in Fig.~\ref{fig:tower_34}.
 The states/operators are classified by the physical symmetry charge $q$. The energy spectrums are rescaled regarding the lowest scaling dimension $\Delta_{\sigma,0}$ as a reference.
At $Q=3$, this rescaling leads to the conformal tower that perfectly matches theoretical values. 
However, when $Q=4$, only lower levels match well due to the strong finite size effect caused by the marginal operator. 

\begin{figure}
\includegraphics[width=.98\textwidth]{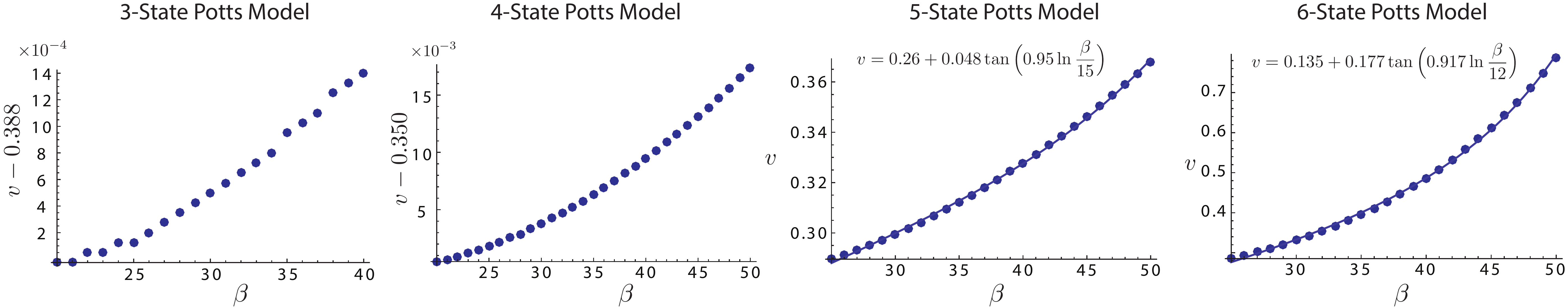}
\caption{Sound velocity of $Q=3$ ($20 \leq \beta \leq 40$ and $\chi=300$), $Q=4$ ($20 \leq \beta \leq 50$ and $\chi=500$), $Q=5$ ($25 \leq \beta \leq 50$ and $\chi=400$) and $Q=6$ ($25 \leq \beta \leq 50$ and $\chi=300$) Potts model at the transition point. \label{fig:v_3456}} 
\end{figure}
The sound velocity can be obtained by $v=\frac{  \beta}{2\pi } \frac{E_{\sigma,0}}{\Delta_{\sigma,0}} $.  
As shown in Fig.~\ref{fig:v_3456}, its dependence on $\beta$ is negligible when $Q=3,4$. We can then estimate $v=0.398435$ for $Q=3$ and $v=0.441824$ for $Q=4$. These values are different from previous studies due to the different algorithm used here.

\begin{figure}[h]
\includegraphics[width=.6\textwidth]{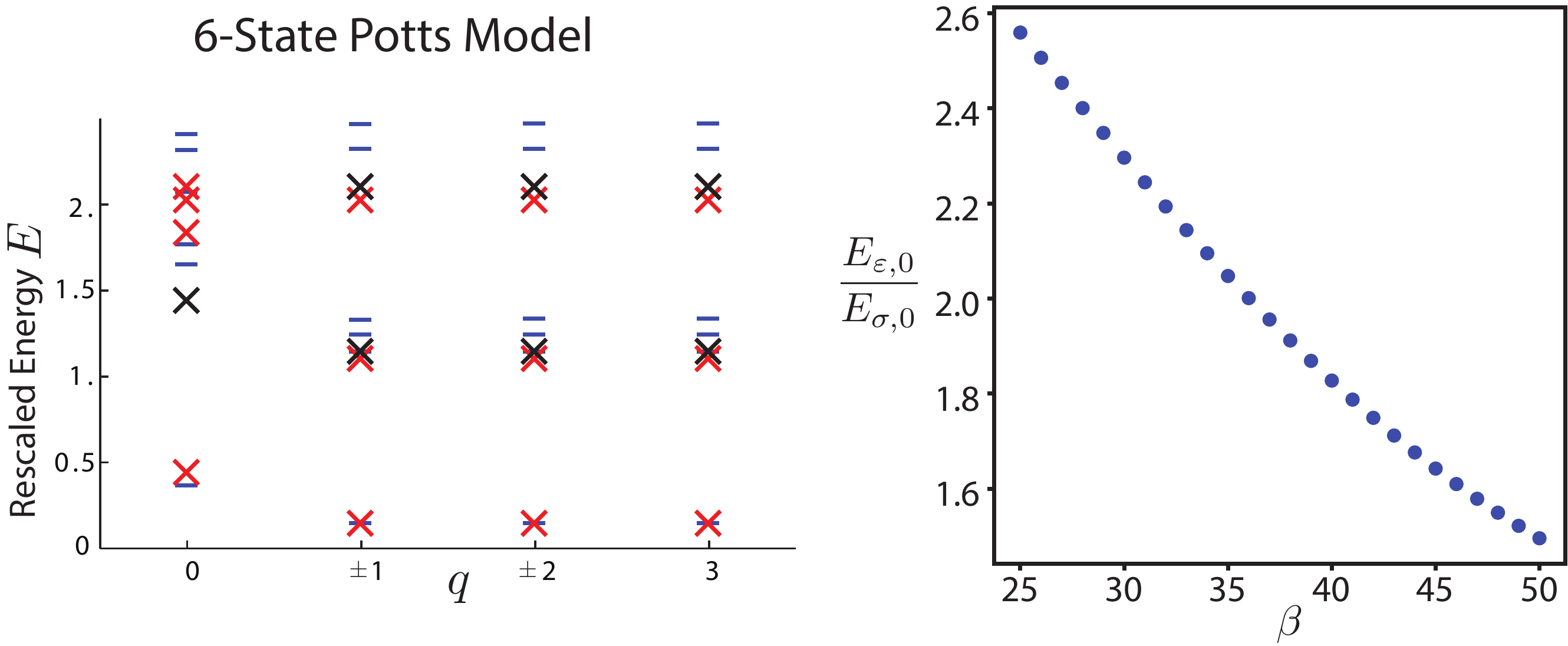}
\caption{(a) Conformal tower of $Q=6$ Potts model at transition point. Red (primary) and Black (descendant) crosses are theoretical values while lines are numerical results. The spectrum is separated into sectors with different symmetry charge (q). (b) The drift behavior of $\Delta_\varepsilon/\Delta_\sigma$. \label{fig:tower_6_drift}}
\end{figure}

Then, we give more numerical study for $Q>4$ Potts model. At $Q=6$, we also obtain the conformal tower by the rescaling regarding $\sigma$ as a reference, as shown in Fig.~\ref{fig:tower_6_drift}(a). Similarly, the theoretical values and numerical data have larger mismatch for higher levels.
To study the drift behavior, we still consider the ratio $\Delta_\varepsilon/\Delta_\sigma$ (shown in Fig.~\ref{fig:tower_6_drift}(b)), which removes the dependence on velocity $v$. This drift at $Q=6$ is more dramatic compared with the drift at $Q=5$ we presented in the main context.

The drift of sound velocity $v$ in this case can be also obtained by $\frac{\beta}{2\pi}E_{\sigma,0}/\Delta_\sigma$. Compare with the drift of $v$ at $Q=3,4$, the drift at $Q=5,6$ (shown in Fig.~\ref{fig:v_3456}) shows much larger $\beta$ dependence.
Inspired by the drift behavior characterized by Eq.~(6), we fit the $\beta$ dependence of $v$ with the formula
\begin{equation}
v = a + b \tan \Big( d \ln \frac{\beta}{\beta_0} \Big)
\end{equation}
for $Q=5,6$, where $a$, $b$, $d$ and $\beta_0$ are parameters to be fitted. As shown in Fig.~\ref{fig:v_3456}, the drift of $v$ becomes severe for larger $Q$, just as the drift behavior of other quantities, e.g., $c$ and $\Delta_\varepsilon/\Delta_\sigma$.

\end{document}